\documentclass[11pt]{article}
\usepackage{amsmath,amssymb, epsfig, graphics}

\newtheorem{theorem}{Theorem}
\newtheorem{lem}{Lemma}[section]

\newcommand{\R}{{\mathbb R}}
\newcommand{\Z}{{\mathbb Z}}

\newcommand{\E}{{\mathbb{E}}}

\def\Z{{\Bbb Z}}
\def\R{{\Bbb R}}

\def\E{{\Bbb E}}

\def\0{{\bf 0}}

\def\phi{\varphi}

\def\T{\T}

\def\Cox{\hfill \Box}

\catcode`@=11 \@addtoreset{equation}{section} \catcode`@=12
\begin{document}
\title{A second row Parking Paradox}

\author{S.R.Fleurke\footnote{Agentschap Telecom, Postbus 450,
9700 AL Groningen, The Netherlands, \texttt{sjoert.fleurke@at-ez.nl}
},\, and C. K\"ulske
\footnote{ University of Groningen, Department of Mathematics and
Computing Sciences, Nijenborg 9, 9747 AC Groningen, The Netherlands,
\texttt{c.kulske@rug.nl}, \texttt{
http://www.math.rug.nl/$\sim$kuelske/ }} }

\maketitle
\begin{abstract}
We consider two variations of the discrete car parking problem
where at every vertex of $\Z$ a car arrives with rate one, now allowing
for parking in two lines. \\
a) The car parks in the first line whenever the vertex and all of
its nearest neighbors are not occupied yet. It can reach the first
line if it is not obstructed by cars already parked in the second
line (``screening'').\\
 b) The car parks according to the same rules,
but parking in the first line can not be obstructed by parked cars
in the second line (``no screening''). In both models, a car that can
not park in the first line will attempt to park in the second line. If
it is obstructed in the second line as well, the attempt is
discarded.\\
We show that both models are solvable in terms of finite-dimensional ODEs.
We compare numerically the limits of first and second line densities, with time going to infinity.
While it is not surprising that model a) exhibits an increase of the
density in the second line from the first line,
more remarkably this is also true for model b), albeit in
a less pronounced way.
\end{abstract}

\vspace{10mm} {\it Key--Words:}
Car parking problem, Random sequential adsorption, Sequential
frequency assignment process, Particle systems.

\section{Introduction}

Car parking, first considered in a mathematical way by R\'{e}nyi \cite{Renyi} in 1958,
gives rise to interesting models that in several variations have been applied in
many fields of science. In the original car parking problem, unit
length cars are appearing with constant rate in time and with constant
density in space on the line where they try to park.
A new car is allowed to park only in case there is no intersection
with previously parked cars.
Otherwise the attempt is rejected. R\'{e}nyi proved that the density
of cars has the limit $0.7475...$, the so-called parking constant.
In the simplest discrete version of the car parking
problem, cars of length $2$ try to park at their midpoints randomly
on $\Z$. This model has been solved analytically as well
\cite{Co62}.
\par
This model belongs to a wider class of more complicated models of
deposition with exclusion interaction. Usually such models are not
analytically solvable. In physical chemistry ``cars'' become
particles which are deposited in layers on a substrate, a process
called random sequential adsorption (RSA). A variety of related
models are studied.  For a review of recent developments see
\cite{Pri07}. Moreover, models with more complicated graphs e.g.
(random) trees have been investigated \cite{Pri04},\cite{Penrose},\cite{Gouet},\cite{Fl08}.

Multilayer variations of the model are used to describe
multilayer adsorption of particles on a substrate \cite{Privman} 
and the sequential frequency assignment process \cite{De07} which
appears in telecommunication. In these papers it is also observed that the
density in higher layers increases up from the first layer, which at first seems 
rather counterintuitive. Heuristic arguments for monotonicity of densities 
were found in specific models \cite{Privman},  but no rigorous proofs could be given yet. 
Moreover Privman  finds numerically a 
scaling behavior of the density  in a similar RSA model \cite{Privman}
with slightly different adhesion rules which is notoriously difficult to explain 
mathematically. 

\par

In the present paper we aim for a rigorous investigation and 
treat two versions of the discrete two-line
car-parking problem with cars of length 2. First we describe the
dynamics of the car parking process without screening and also with
screening. Then  we provide the {\em solutions} of these models by
reducing them to closed finite dimensional systems of ODEs for
densities of local patterns, see Theorems \ref{Theo1} and
\ref{Theo2}. That it is possible to find a finite-dimensional dynamical
description is quite remarkable. It is not obvious, and in fact our
method ceases to work for a three-line extension of the model
without screening where an infinite system appears.

A second remarkable fact is that, even without screening, the second
line density is higher than the first. Cars do not communicate or
plan a common strategy and their arrival is random, but they seem to
use the resources in the second line more efficiently, once they
have been rejected in the first line.
%
\section{The Dynamics}

We will define a Markov jump process on the (suitably coded)
occupation numbers $m=(m_i)_{i\in \Z}\in \Omega=\{0,1,2,3\}^\Z$.

Here the {\it spin} $m_i$ denotes the joint occupation numbers at
vertex $i$ at height $1$ and $2$.
 It is useful for short notation to interpret the occupation numbers
 at various heights as binary digits and write ordinary
 natural numbers. That is we write
 \[ m_{i} = \left\{ \begin{array}{rl}
0 & \mbox{if vertex $i$ is vacant in the first and second line} \\
1 & \mbox{if vertex $i$ is occupied in the first but not in the second line} \\
2 & \mbox{if vertex $i$ is occupied in the second but not in the first line} \\
3 & \mbox{if vertex $i$ is occupied in the first and in the second
line}
  \end{array} \right. \]
so that $m_i\in \{0,1,2,3\}$. The dynamics of the process is defined
in terms of the generator which is given by the right hand side of
the differential equation
\begin{equation}\begin{split}
\frac{d}{dt} \mathbb{E}^{m} f(m(t)) = \sum_{k,s} \left[ f(m^{s,k}) -
f(m) \right] c(s;m_{k-1},m_{k},m_{k+1}) \label{Generator}
\end{split} \end{equation}
with
\[ m^{s,k}_i = \left\{ \begin{array}{rl}
m_i & \mbox{if $k \neq i$ } \\
s & \mbox{if $k = i$ }  \end{array} \right. \] denoting the
configuration which has been obtained by $m$ by changing the
configuration in $i$ to $s$. Here $\E^{m}$ denotes the expected
value with respect to the process, started at the initial
configuration $m$.

\subsubsection*{Two-line parking rates}

The rates are either equal to zero or one. They are $1$ precisely in
the following cases.
\begin{enumerate}
\item{$0\mapsto 1$}
Adding a car in the first line at site $i$. For the model without
screening we have
\begin{equation}\begin{split}
c(1;0,0,0)&=c(1;2,0,0)=c(1;0,0,2)=c(1;2,0,2)=1\cr
\end{split} \end{equation}
Indeed, this occurs when the site itself is empty on the first and
second line and the nearest neighbors are empty in the first line,
see figure \ref{fig:01} for an example.
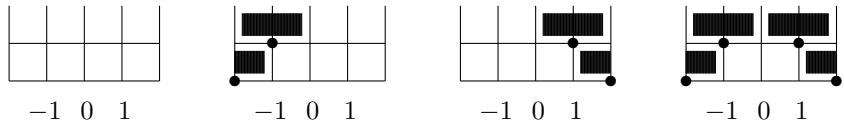
\begin{figure}[h]
\setlength{\unitlength}{1mm}
\begin{picture}(150,30)(0,-7)

%
\multiput(20,5)(30,0){4}{\multiput(0,0)(5,0){5}{\line(0,1){10}}}
\multiput(20,5)(30,0){4}{\line(1,0){20}}
\multiput(20,10)(30,0){4}{\line(1,0){20}}
\put(55,10){\circle*{1.5}} \put(50,5){\circle*{1.5}}
\put(95,10){\circle*{1.5}} \put(100,5){\circle*{1.5}}
\put(115,10){\circle*{1.5}} \put(110,5){\circle*{1.5}}
\put(125,10){\circle*{1.5}} \put(130,5){\circle*{1.5}}
\multiput(51,11)(.1,0){80}{\line(0,1){3}}
\multiput(50,6)(.1,0){40}{\line(0,1){3}}
\multiput(91,11)(.1,0){80}{\line(0,1){3}}
\multiput(96,6)(.1,0){40}{\line(0,1){3}}
\multiput(111,11)(.1,0){80}{\line(0,1){3}}
\multiput(110,6)(.1,0){40}{\line(0,1){3}}
\multiput(121,11)(.1,0){80}{\line(0,1){3}}
\multiput(126,6)(.1,0){40}{\line(0,1){3}}
\multiput(22.5,0)(30,0){4}{\put(0,0){\small{$-1$}}}
\multiput(29.5,0)(30,0){4}{\put(0,0){\small{$0$}}}
\multiput(34.5,0)(30,0){4}{\put(0,0){\small{$1$}}}

\end{picture} \caption{Configurations of vertices $-1, 0,$ and $1$ that allow a transition from $m_0=0$ to $m_0=1$
in the model without screening. In the model with screening only the
most left configuration allows a transition to $m_0=1$. }
\label{fig:01}
\end{figure}

In the screening model however, cars in the second line will
obstruct cars from reaching the first line. Therefore, in the
screening model we have as the only nonvanishing rate
\begin{equation}\begin{split}
c^{\hbox{sc}}(1;0,0,0)&=1\cr
\end{split} \end{equation}

\item{$0\mapsto 2$} Adding a car in the second line at $i$ while the first line was empty at the site
\begin{equation}\begin{split}
c(2;1,0,0)&=c(2;0,0,1)=c(2;1,0,1)=1\cr
\end{split} \end{equation}
Indeed, this occurs when there was a supporting site $i-1$ or $i+1$
or both with one car in the first line. This is true for both
models.

\item{$1\mapsto 3$} Adding a car in the second line while the first line was full at the site
\begin{equation}\begin{split}
c(2;0,1,0)&=1\cr
\end{split} \end{equation}
Indeed, this occurs when there are no obstructing cars right and
left at height $2$. There can be no obstructing cars right and left
at height $1$ because there could not be a car in the first line at
$i$ otherwise. This is true for both models.
\end{enumerate}
All other transitions are impossible.
\par
This generator defines a Markov jump process on the infinite graph
by standard theory \cite{Liggett}, such that  (\ref{Generator})
holds for any local function $f: \Omega\rightarrow\R$.

\section{Results}

We provide a closed system of differential equations for the
densities of occupied sites, involving densities of finitely many local patterns,
in both models. First we need some
definitions. Here and in the following we use for the densities at single sites,
and triples of sites the notation
\begin{equation}\begin{split}
D_t(s)&:=P_t(m_{0}=s)\cr
D_t(s_{-1},s_{0},s_{1})&:=P_t(m_{-1}=s_{-1},m_{0}=s_{0},m_{1}=s_{1})\cr
\label{def:Dt}
\end{split}\end{equation}
Further we need the following ``one-sided densities''
\begin{equation}\begin{split}
f_t (s)&:=P_t(m_{1}=s | N_0(t)=0), \quad \text{   for } s=0,1,2\cr
R_t &:=P_t(m_{1}=1, m_{2}=0 |
N_0(t)=0)
\label{osq}
\end{split}\end{equation}
where $N_j(t)$ denotes the Poisson counting process of events of car
arrivals at site $j$.

As our main result we show that the time-evolution of these
densities gives rise to a closed ODE.

\begin{theorem}
\label{Theo1}
Two-line Parking without Screening. \\ The time evolution of the probability vector
$(D_t(0),D_t(1),D_t(2),D_t(3))$ obeys the following system of
differential equations.
\begin{equation}\begin{split}
\frac{d}{dt}D_t(0)&= -(f_t(0) + f_t(2))^2e^{-t} - (2f_t(0)f_t(1)
+f_t(1)^2)e^{-t}\cr %
\frac{d}{dt}D_t(1)&= (f_t(0) + f_t(2))^2 e^{-t} - D_t(0,1,0) \cr %
\frac{d}{dt}D_t(2)&= (2f_t(0)f_t(1)
+f_t(1)^2)e^{-t}\cr %
\frac{d}{dt}D_t(3)&= D_t(0,1,0) \cr %
\end{split}\end{equation}
with initial conditions $D_{0}(s)=1_{s=0}$, where the vector
$(f_t(0),f_t(1),f_t(2),R_t)$  obeys the linear ODE
\begin{equation}\begin{split}\label{osqODE}
\frac{d}{dt}f_t(0)&= - f_t(0) e^{-t }  - f_t(1) e^{-t }- f_t(2)e^{-t}\cr%
\frac{d}{dt}f_t(1)&= f_t(0)e^{-t} +f_t(2)e^{-t}- R_t\cr%
\frac{d}{dt}f_t(2)&= f_t(1)e^{-t}\cr %
\frac{d}{dt}R_t&= f_t(0)(e^{-t}-t e^{-2 t })-f_t(1) t e^{-2 t } -R_t
\end{split}\end{equation}
with initial conditions $(f_0(0),f_0(1),f_0(2),R_0)=(1,0,0,0)$, \\
and finally, $D_t(0,1,0)$ is obeying the equation
\begin{equation}\begin{split}
\frac{d}{dt}D_t(0,1,0)&= f_t(0)^2 e^{-t}- D_t(0,1,0) -2 R_t  f_t(0)
e^{-t}-2 R_t f_t(1)e^{-t}\cr
\end{split}\end{equation}
with $D_0(0,1,0)=0$.
\end{theorem}

\begin{figure}[htp]
\includegraphics[height=5.5cm]{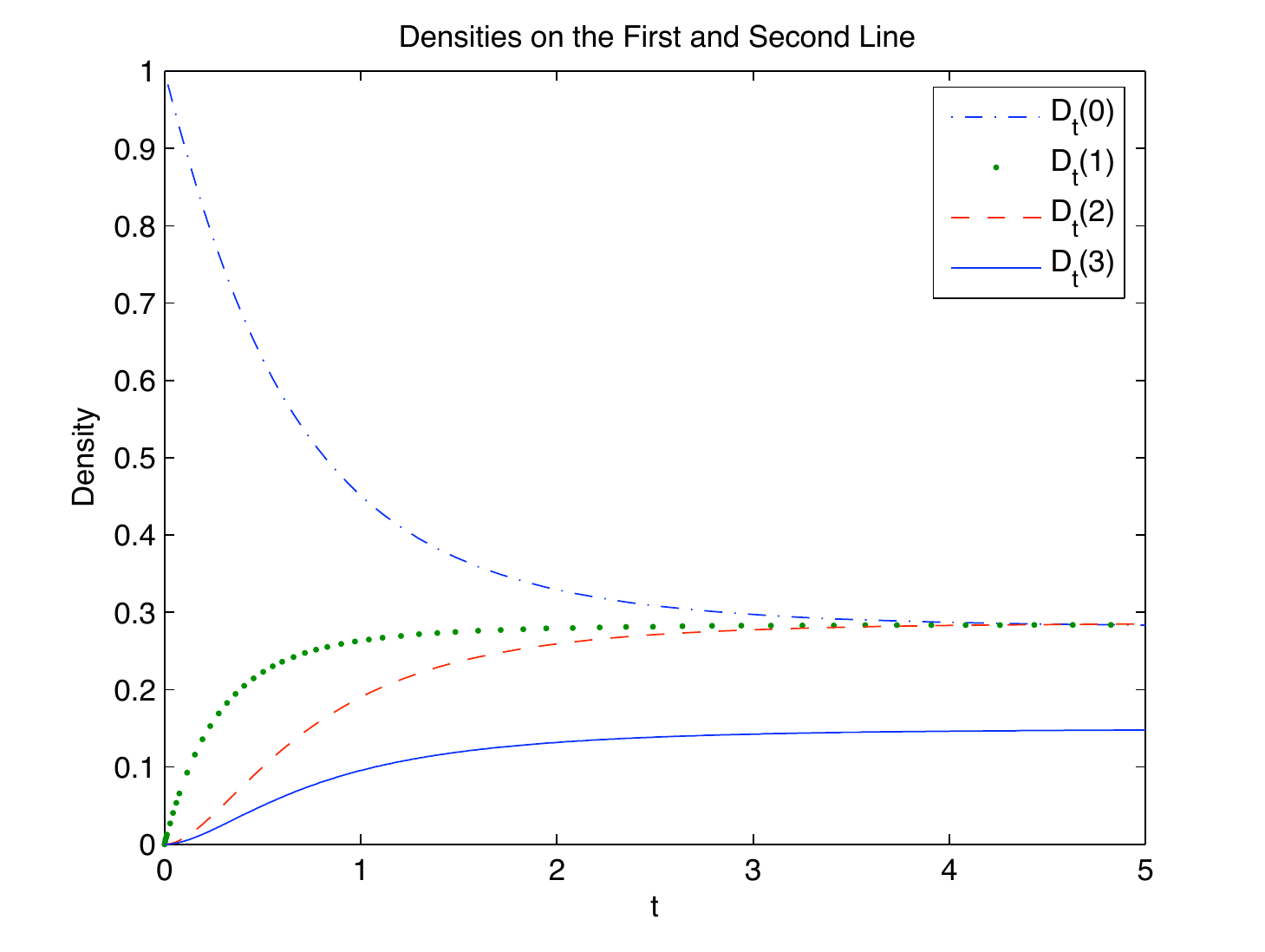}
\includegraphics[height=5.5cm]{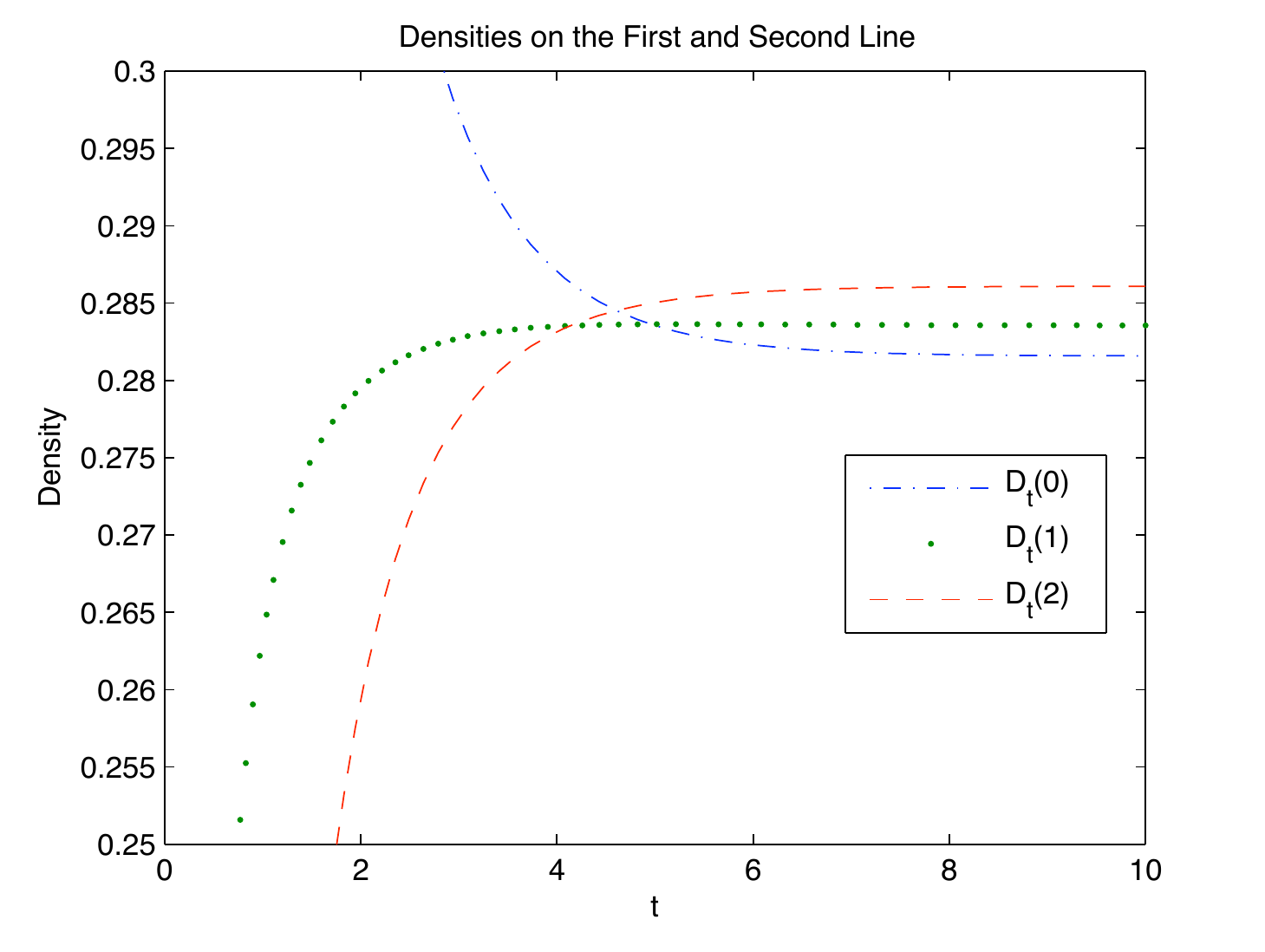}
$\phantom{123}$

 \caption{Numerical solution of the densities in the first and second line. The right figure
  zooms in on the limits of $D_t(1)$ and $D_t(2)$ to show they tend to different values.
 }
  \label{fig:D}
\end{figure}

The system above can be solved numerically and the results are
depicted in figure \ref{fig:D}. As it can be seen in the right
figure, surprisingly the value of $D_t(2)$ has a slightly higher
limit than $D_t(1)$. This clearly means that the second line has a
higher limit density of cars than the first line. This result is
independently confirmed by simulations of the parking process
measuring the empirical densities.

A similar system of equations holds for the model with screening.
Recall that in this model cars are not allowed to pass cars on the
second line to reach a void on the first line. This results in less
possibilities of filling voids of the first line than in the model
treated above. In fact we can derive the ODEs of this model by
simply deleting those terms in (\ref{osqODE}) that represent the
possibility of ``jumping'' over a car in the second line to reach a
void on the first line. So, we get


\begin{theorem}\label{Theo2} Two-line Parking with Screening\\
The time evolution of the probability vector $(D_t^{sc}(0),
D_t^{sc}(1), D_t^{sc}(2), D_t^{sc}(3))$ obeys the following system
of differential equations
\begin{equation}\begin{split}
\frac{d}{dt}D_t^{sc}(0)&= -(f_t^{sc}(0) + f_t^{sc}(1))^2 e^{-t}\cr %
\frac{d}{dt}D_t^{sc}(1)&= f_t^{sc}(0)^2 e^{-t} - D_t^{sc}(0,1,0) \cr %
\frac{d}{dt}D_t^{sc}(2)&= (2f_t^{sc}(0)f_t^{sc}(1)
+f_t^{sc}(1)^2)e^{-t}\cr %
\frac{d}{dt}D_t^{sc}(3)&= D_t^{sc}(0,1,0) \cr %
\end{split}\end{equation}
with initial conditions $D_0^{sc}(s)=1_{s=0}$, where the vector
$(f_t ^{sc}(0), f_t ^{sc}(1), f_t ^{sc}(2), R_t ^{sc})$ obeys the
linear ODE
\begin{equation}
\begin{split}
\frac{d}{dt}f_t^{sc}(0)&= - f_t^{sc}(0) e^{-t }  - f_t^{sc}(1) e^{-t }\cr%
\frac{d}{dt}f_t^{sc}(1)&= f_t^{sc}(0)e^{-t} - R_t^{sc}\cr%
\frac{d}{dt}f_t^{sc}(2)&= f_t^{sc}(1)e^{-t}\cr %
\frac{d}{dt}R_t^{sc}   &= f_t^{sc}(0)(e^{-t}-t e^{-2 t})-f_t^{sc}(1)
t e^{-2 t } -R_t^{sc} \cr
\end{split}
\end{equation}
with initial conditions $(f_0^{sc}(0),f_0^{sc}(1),f_0^{sc}(2),R_0^{sc})=(1,0,0,0)$, \\
and finally, $D_t^{sc}(0,1,0)$ is obeying the equation
\begin{equation}\begin{split}
\frac{d}{dt}D_t^{sc}(0,1,0)&= f_t^{sc}(0)^2 e^{-t}- D_t^{sc}(0,1,0)
-2 R_t^{sc} f_t^{sc}(0) e^{-t}-2 R_t^{sc} f_t^{sc}(1)e^{-t}\cr
\end{split}\end{equation}
with $D_0^{sc}(0,1,0)=0$.
\end{theorem}

\section{Proofs of Theorem 1 and Theorem 2}

The following lemmas are used to prove our theorems.
\begin{lem}
\label{D}
The probability vector $(D_t(0), D_t(1), D_t(2), D_t(3))$
obeys
\begin{eqnarray}
\frac{d}{dt}D_t(0)&=& -D_t(0,0,0) -2 D_t(2,0,0)-D_t(2,0,2)-2
D_t(1,0,0)-D_t(1,0,1) \label{D0} \\
\frac{d}{dt}D_t(1)&=& D_t(0,0,0) +2 D_t(2,0,0)+D_t(2,0,2)-D_t(0,1,0)
\label{D1}\\
\frac{d}{dt}D_t(2)&=& 2 D_t(1,0,0)+D_t(1,0,1) \\
\frac{d}{dt}D_t(3)&=& D_t(0,1,0)
\end{eqnarray}
\end{lem}

{\bf Remark: } Summing over the four right hand sides we get zero,
due to the fact that we have summed a probability vector. It is also
interesting to check that
\begin{equation}\begin{split}
\frac{d}{dt}D_t(1)+\frac{d}{dt}D_t(3)&= D_t(0,0,0) +2
D_t(2,0,0)+D_t(2,0,2)\cr
\end{split}\end{equation}
recovers the ODE for the density in the first line.\\

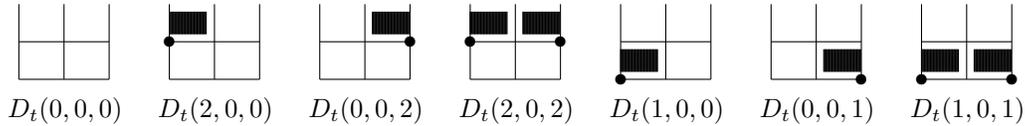
\begin{figure}[h]
\setlength{\unitlength}{1mm}
\begin{picture}(150,30)(0,-7)

%
\multiput(10,5)(20,0){7}{\line(0,1){10}}
\multiput(10,10)(20,0){7}{\line(1,0){12}}
\multiput(16,5)(20,0){7}{\line(0,1){10}}
\multiput(22,5)(20,0){7}{\line(0,1){10}}
\multiput(10,5)(20,0){7}{\line(1,0){12}} %
%
\put(30,10){\circle*{1.5}} \put(62,10){\circle*{1.5}}
\put(70,10){\circle*{1.5}} \put(82,10){\circle*{1.5}}
\put(90,5){\circle*{1.5}} \put(122,5){\circle*{1.5}}
\put(130,5){\circle*{1.5}} \put(142,5){\circle*{1.5}}
\put(8.5,0){\small{$D_t(0,0,0)$}} \put(28.5,0){\small{$D_t(2,0,0)$}}
\put(48.5,0){\small{$D_t(0,0,2)$}}
\put(68.5,0){\small{$D_t(2,0,2)$}}
\put(88.5,0){\small{$D_t(1,0,0)$}}
\put(108.5,0){\small{$D_t(0,0,1)$}}
\put(128.5,0){\small{$D_t(1,0,1)$}}

\multiput(30,11)(.1,0){50}{\line(0,1){3}}
\multiput(57,11)(.1,0){50}{\line(0,1){3}}
\multiput(70,11)(.1,0){50}{\line(0,1){3}}
\multiput(77,11)(.1,0){50}{\line(0,1){3}}
\multiput(90,6)(.1,0){50}{\line(0,1){3}}
\multiput(117,6)(.1,0){50}{\line(0,1){3}}
\multiput(130,6)(.1,0){50}{\line(0,1){3}}
\multiput(137,6)(.1,0){50}{\line(0,1){3}}

\end{picture} \caption{List of all occupancy configurations of vertices -1, 0 and 1 in the first and second line
that (may) contribute to a reduction of the proportion of $D_t(0)$.}
\label{fig:D0}
\end{figure}

{\bf Proof: } Fix an arbitrary vertex. Let us call this vertex $0$.
Starting from the dynamics (\ref{Generator}) and using symmetries we
have
\begin{equation}\begin{split}
\frac{d}{dt}D_t(0)&=-D_t(0,0,0) -2 D_t(2,0,0)-D_t(2,0,2)-2
D_t(1,0,0)-D_t(1,0,1) \cr
\end{split}\end{equation}
Indeed, the first three terms correspond to adding a car in the
first line, the next two terms correspond to adding a car in the
second line, see figure \ref{fig:D0}.

The other three differential equations are derived in a similar way.

\begin{lem} The triple site densities $D_t(s,0,s')$ and the one-sided densities as defined in
\ref{def:Dt} and \ref{osq} respectively are related in the following
way
\begin{equation}\begin{split}
D_t(s,0,s')&= f_t(s)f_t(s') e^{- t} \cr
\end{split}\end{equation}
for $(s,s')\in \{(0,0),(0,1),(1,0),(1,1),(2,2)\}$. \label{D&osq}
\end{lem}

{\bf Proof: } We note that for the mentioned choices of $(s,s')$
conditioning on non-arrival at zero does not change the probability,
that is
\begin{equation}\begin{split}
D_t(s,0,s')&= P_t(m_{-1}=s,m_{0}=0,m_{1}=s' , N_0(t)=0)\cr &=
P_t(m_{-1}=s,m_{0}=0,m_{1}=s' | N_0(t)=0) e^{- t} \cr
\end{split}\end{equation}

In the next step we note that, conditional on the event that no car
has arrived at the site $0$, the dynamics for the two sides that are
emerging from $0$ is independent. Consequently we have
\begin{equation}\begin{split}\label{2.7}
&P_t(m_{-1}=s,m_{0}=0,m_{1}=s' | N_0(t)=0)\cr &= P_t(m_{1}=s |
N_0(t)=0)P_t(m_{1}=s' | N_0(t)=0)\cr \cr
\end{split}\end{equation}
This concludes the proof of the Lemma.
$\Cox$

Next we look at the time-evolution of the ``one-sided densities''.

\begin{lem}
\label{lem:osq} %
The vector $(f_t(0),f_t(1),f_t(2),R_t)$ obeys the ODE
\begin{equation}\begin{split}\label{2.7a}
\frac{d}{dt}f_t(0)&= - f_t(0) e^{-t }  - f_t(1) e^{-t }- f_t(2)e^{-t}\cr%
\frac{d}{dt}f_t(1)&= f_t(0)e^{-t} +f_t(2)e^{-t}- R_t\cr%
\frac{d}{dt}f_t(2)&= f_t(1)e^{-t}\cr %
\frac{d}{dt}R_t&= f_t(0)(e^{-t}-t e^{-2 t })-f_t(1) t e^{-2 t } -R_t
\end{split}\end{equation}
with initial conditions $f_{t=0}(s)=1_{s=0}$ and $R_{t=0}=0$.

\end{lem}

{\bf Remark 1: } Note that combining the equations of
$\frac{d}{dt}f_t(0)$ and $\frac{d}{dt}f_t(2)$ readily gives
\begin{equation}
f_t(0) + f_t(2) = \exp(e^{-t}-1)
\end{equation}
which is a known result for the first line in a semi-infinite chain
\cite{Co62}.

{\bf Remark 2: } Note also that because $R_t := P_t(m_{1}=1, m_{2}=0
| N_0(t)=0)$ we have in fact $R_t = \frac{d}{dt} f_t(3)$.


{\bf Proof: } To derive ODEs for these densities we employ the
generator of the process, while putting the term at the site $0$ to
sleep, and correspondingly the spin at zero to be the constant
$m_0=0$. %
For the first quantity we get
\begin{equation}\begin{split}\label{2.7b}
\frac{d}{dt}f_t(0)&=\frac{d}{dt}P_t(m_{1}=0 | N_0(t)=0)\cr %
&= -P_t(m_{1}=0 , m_{2}=0| N_0(t)=0) - P_t(m_{1}=0 ,m_{2}=2|N_0(t)=0) \cr%
&~- P_t(m_{1}=0 , m_{2}=1| N_0(t)=0)\cr %
&= - P_t(m_{1}=0 , m_{2}=0|N_0(t)=0, N_1(t)=0)e^{-t} \cr%
&~-P_t(m_{1}=0 , m_{2}=2| N_0(t)=0, N_1(t)=0)e^{-t} \cr
&-P_t(m_{1}=1| N_0(t)=0)e^{-t }\cr %
&= - P_t(m_{1}=0| N_0(t)=0)e^{-t} - P_t(m_{1}=2 | N_0(t)=0)e^{-t}\cr%
&- P_t(m_{1}=1| N_0(t)=0)e^{-t }\cr &= - f_t(0) e^{-t }  - f_t(1)
e^{-t }- f_t(2)e^{-t }\cr \cr
\end{split}\end{equation}
Next we have
\begin{equation}\begin{split}\label{2.7c}
\frac{d}{dt}f_t(1)&= \frac{d}{dt}P_t(m_{1}=1 | N_0(t)=0)\cr &=
P_t(m_{1}=0 , m_{2}=0| N_0(t)=0) +P_t(m_{1}=0 , m_{2}=2|
N_0(t)=0)\cr &- P_t(m_{1}=1 , m_{2}=0| N_0(t)=0)\cr &= f_t(0)e^{-t}
+f_t(2)e^{-t} - R_t\cr \cr
\end{split}\end{equation}
and
\begin{equation}\begin{split}\label{2.7d}
\frac{d}{dt}f_t(2)&= \frac{d}{dt}P_t(m_{1}=2 | N_0(t)=0)\cr &=
P_t(m_{1}=0 , m_{2}=1| N_0(t)=0)\cr &= P_t(m_{1}=1|
N_0(t)=0)e^{-t}\cr
\end{split}\end{equation}
Finally, we get
\begin{equation}\begin{split}
\frac{d}{dt}R_t&= \frac{d}{dt}P_t(m_{1}=1, m_{2}=0 | N_0(t)=0) \cr %
&= P_t(m_{1}=0, m_{2}=0 | N_0(t)=0) -P_t(m_{1}=1, m_{2}=0 |
N_0(t)=0)\cr &-P_t(m_{1}=1, m_{2}=0 , m_3=0| N_0(t)=0) \cr%
&-P_t(m_{1}=1, m_{2}=0 , m_3=1| N_0(t)=0)
\end{split}\end{equation}
Using conditioning on non-arrival again we get
\begin{equation}\begin{split}
&\frac{d}{dt}P_t(m_{1}=1, m_{2}=0 | N_0(t)=0)\cr &= P_t(m_{1}=0|
N_0(t)=0)e^{-t } -P_t(m_{1}=1, m_{2}=0 | N_0(t)=0)\cr &-P_t(m_{1}=1|
N_0(t)=0,N_2(t)=0 )P_t(m_{1}=0| N_0(t)=0)e^{-t }\cr &-P_t(m_{1}=1|
N_0(t)=0,N_2(t)=0 )P_t(m_{1}=1| N_0(t)=0)e^{-t } \cr
\end{split}\end{equation}
Clearly we have
\begin{equation}\begin{split}
&P_t(m_{1}=1| N_0(t)=0,N_2(t)=0 )= t e^{-t}
\end{split}\end{equation}
because there is precisely one car at $1$ if and only if precisely
one car arrived conditioning on no cars at $0$ and $2$. This shows
that the last ODE is correct and concludes the proof of the lemma.$\Cox$

The only remaining term whose time-evolution we need to consider is $D_t(0,1,0)$.

\begin{lem} $D_t(0,1,0)$ is a solution of the
differential equation

\begin{equation}\begin{split}
\frac{d}{dt}D_t(0,1,0)&= f_t(0)^2 e^{-t}- D_t(0,1,0) -2 R_t  f_t(0)
e^{-t}-2 R_t f_t(1)e^{-t}\cr
\end{split}\end{equation}
\label{D010}
\end{lem}

{\bf Proof: } We note that
\begin{equation}\begin{split}
\frac{d}{dt}D_t(0,1,0)&= D_t(0,0,0) - D_t(0,1,0)-2 D_t(0,1,0,0)- 2
D_t(0,1,0,1)\cr \cr
\end{split}\end{equation}
The first term is for adding a car at the central site from the
vacuum, the second for adding a car at the central site at height
one. The last two terms are for adding a car to the right 
of the central site. As we already know we have
\begin{equation}\begin{split}
D_t(0,0,0) = P_t(m_{1}=0 | N_0(t)=0)^2 e^{-t} \cr
\end{split}\end{equation}
Using conditioning on non-arrival at $2$ we get, by reflection
invariance
\begin{equation}\begin{split}
D_t(0,1,0,0) &= P_t(m_{1}=1, m_{2}=0 | N_0(t)=0)  P_t(m_{1}=0 |
N_0(t)=0) e^{-t} \cr &= R_t f_t(0) e^{-t}
\end{split}\end{equation}
For the last term we get in the same way
\begin{equation}\begin{split}
D_t(0,1,0,1) &= P_t(m_{1}=1, m_{2}=0 | N_0(t)=0)  P_t(m_{1}=1 |
N_0(t)=0) e^{-t} \cr &= R_t f_t(1) e^{-t}\cr
\end{split}\end{equation}
$\Cox$\\

{\bf Proof of Theorem \ref{Theo1}: } Combining the results of lemma
\ref{D}, \ref{D&osq}, \ref{lem:osq} and \ref{D010} proves Theorem 1. $\Cox$ \\

{\bf Proof of Theorem \ref{Theo2}: } The proof follows from Theorem
\ref{Theo1} by deleting every term that represents the possibility
of skipping a second line car to reach a void in the first line.
This results in deleting $f_t(2)e^{-t}$ from the first two equations
of \ref{lem:osq}, and $D_t(2,0,0)$ and $D_t(2,0,2)$ from \ref{D0}
and \ref{D1}. $\Cox$

\begin{figure}[htp]
\includegraphics[height=5.5cm]{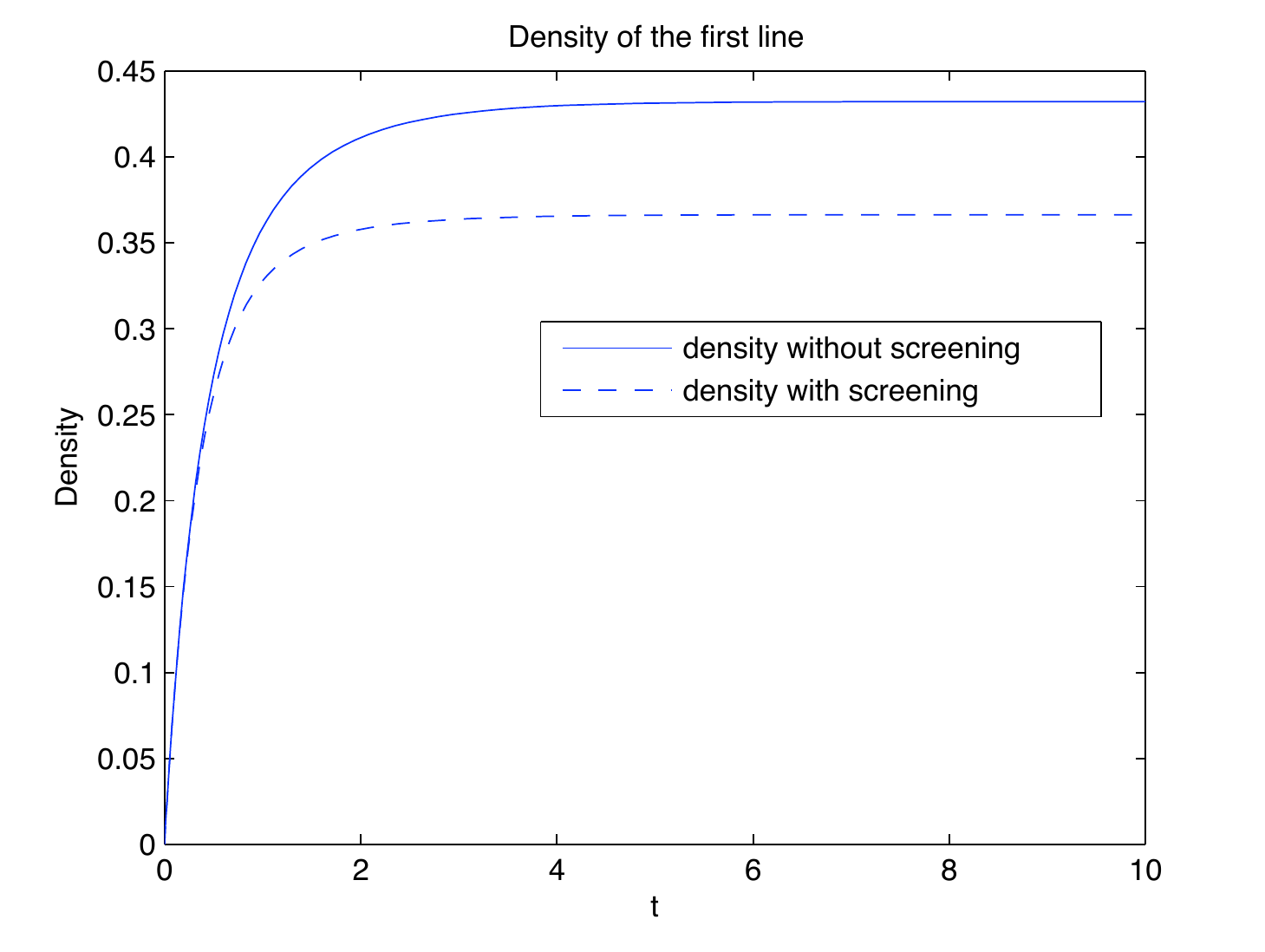}
\includegraphics[height=5.5cm]{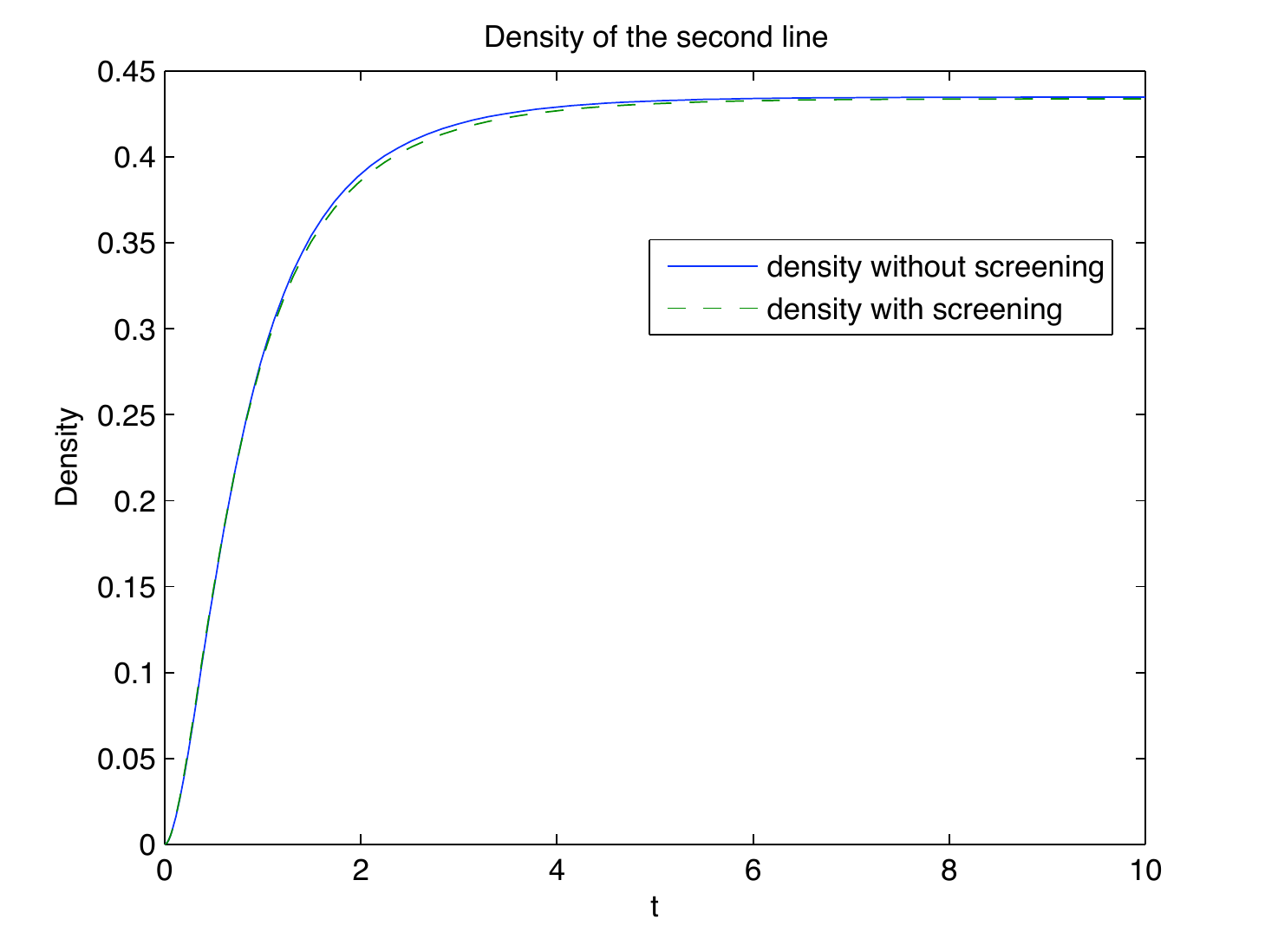}
$\phantom{123}$

  \hfill
  \caption{Densities of the first and second line.} \label{fig:DD}
\end{figure}

\section{Conclusion}
We introduced two extensions of the classic parking problem to a
two-line model i.e. a model with screening and a model without
screening. For both models we derived closed systems of finite-dimensional 
ODEs from which the time-evolution of the densities in the first and second line 
can be obtained.   
Interestingly, the numerical solution of the ODE shows that the final densities in the second line
are higher than those in the first line, for both models. The increase factor in the
model without screening is approximately
\[ I = \lim_{t \rightarrow \infty} \frac{D_t(2) + D_t(3)}{D_t(1) + D_t(3)}
\approx \frac{0.434868 }{0.432332 } \approx 1.006 \]
It is known by analytical computations  \cite{Co62} that  $D_t(1) + D_t(3)$ approaches $(1-e^{-2})/2\approx 0.432332$ for large $t$,
which provides a checkup for the numerics.
In the screening model we find
\[ I_{sc} = \lim_{t \rightarrow \infty} \frac{D_t^{sc}(2) + D_t^{sc}(3)}{D_t^{sc}(1) + D_t^{sc}(3)}
\approx \frac{0.433896}{0.366475} \approx
1.184 \] %

In other words, in both models the cars seem to exploit the resources in the second line in a 
(slightly) more efficient way than in the first line.

\section*{Acknowledgements}  
The authors thank Aernout van Enter and Herold Dehling for interesting discussions.

\end{document}